\documentclass[a4paper,11pt]{article}
\pdfoutput=1
\usepackage{jheppub}
\usepackage{amsmath,amssymb,latexsym}
\usepackage{braket}
\usepackage{bm}
\usepackage{booktabs}
\usepackage{setspace}
\usepackage{hepunits}
\usepackage[svgnames]{xcolor}
\usepackage{graphicx}

\usepackage{mciteplus}

\newcommand{\ep}{\varepsilon}
\newcommand{\nn}{\nonumber}
\newcommand{\nd}{\mathrm{d}}

\allowdisplaybreaks

\def\eqref#1{(\ref{#1})}
\newcommand{\bea}{\begin{eqnarray}}
	\newcommand{\eea}{\end{eqnarray}}
\newcommand{\bean}{\begin{eqnarray*}}
	\newcommand{\eean}{\end{eqnarray*}}

\title{
Towards Systematic Evaluation of de Sitter Correlators via  Generalized Integration-By-Parts
Relations
}

\author[a]{Jiaqi Chen}
\author[a,b]{Bo Feng}

\affiliation[a]{Beijing Computational Science Research Center, Beijing 100084, China}
\affiliation[b]{Peng Huanwu Center for Fundamental Theory, Hefei, Anhui 230026, China}

\emailAdd{jiaqichen@csrc.ac.cn}
\emailAdd{fengbo@csrc.ac.cn}

\abstract{We generalize Integration-By-Parts (IBP) and differential equations methods to  de Sitter correlators related to inflation. While massive correlators in de Sitter spacetime  are usually regarded as highly intricate, we find they have remarkably hidden concise structures from the perspective of IBP. We find the factorization of the IBP relations of each vertex integral family corresponding to $\nd \tau_i$ integration. Furthermore, with a smart construction of master integrals, the universal formulas for iterative reduction and $\nd \log$-form differential equations of arbitrary vertex integral family are presented and proved. These formulas dominate all tree-level de Sitter correlators and play a kernel role at the loop-level as well.}

\begin{document}

\maketitle


\section{Introduction}

The  quantum field theory in curved spacetime has a rich history and holds paramount significance in various physical applications. As the maximally symmetric spacetime, de Sitter (dS) spacetime as well as anti-de Sitter (AdS) spacetime should be the simplest  curved spacetime to study. The quantum field theory in dS and AdS  has gained large interest among researchers. In the phenomenology of particle physics and cosmology, correlators in dS spacetime play a crucial role in the study of inflationary cosmologies and possible signals of new physics, known as the cosmological
collider program \cite{Chen:2009we,Chen:2009zp,Arkani-Hamed:2015bza}. These correlators are then related to experimental observations \cite{Achucarro:2022qrl} such as Cosmic Microwave Background (CMB) radiation and Large-Scale Structure (LSS). Utilizing the in-in formalism \cite{schwinger1961brownian} in cosmology  \cite{Calzetta:1986ey,Weinberg:2005vy} and its diagrammatic approach, the Schwinger-Keldysh (SK) path integral \cite{schwinger1961brownian,keldysh1965diagram,feynman1963theory}, one can derive Feynman-like diagrams and rules for  $n$-point correlators \cite{Chen:2017ryl}. The computation and analysis of these inflation correlators is an ongoing research topic. People have considered this problem from many aspects such as cosmological bootstrap \cite{Arkani-Hamed:2018kmz,Baumann:2019oyu,Baumann:2020dch,Pajer:2020wnj,Hillman:2021bnk,Baumann:2021fxj,Hogervorst:2021uvp,Pimentel:2022fsc,Jazayeri:2022kjy,Wang:2022eop,Baumann:2022jpr}, cutting rule \cite{Goodhew:2020hob,Goodhew:2021oqg,Melville:2021lst,DiPietro:2021sjt,Tong:2021wai,Salcedo:2022aal,Agui-Salcedo:2023wlq}, Mellin-Barnes (MB) space \cite{
Sleight:2019hfp,Sleight:2019mgd,Sleight:2020obc,Sleight:2021plv,Jazayeri:2021fvk,Premkumar:2021mlz}, partial MB transformation \cite{Qin:2022lva,Qin:2022fbv,Qin:2023ejc,Qin:2023bjk}, integrand issues \cite{Xianyu:2022jwk,Loparco:2023rug, Arkani-Hamed:2017fdk,Arkani-Hamed:2018bjr,Lee:2022fgr,Gomez:2021qfd,Gomez:2021ujt}. Nevertheless, this topic, to the best of our current knowledge, has not been systematically solved and comprehended.

On the other side, the methods developed for computing amplitudes in flat spacetime are well-established in last two decades. Similar methods have recently been applied
for dS correlators.
For example, MB transformation has been applied in both flat \cite{Smirnov:1999gc,Tausk:1999vh} and dS \cite{Sleight:2019hfp,Sleight:2019mgd,Sleight:2020obc,Sleight:2021plv,Jazayeri:2021fvk,Premkumar:2021mlz, Qin:2022lva,Qin:2022fbv,Qin:2023ejc,Qin:2023bjk}. 
In dS case, people find the (high-order) differential equation of correlator and  solve it in terms of hypergeometric functions \cite{Qin:2023ejc}. Meanwhile, in flat space, similar method also has been considered, using the more systematic mathematical tool GKZ \cite{Kalmykov:2012rr,Ananthanarayan:2022ntm,beukers2016monodromy,delaCruz:2019skx,Klausen:2019hrg}. 
Thus, examining the distinct features of amplitudes in flat spacetime and correlators in dS spacetime and drawing insights from the  computation methods for flat spacetime could provide guidance for developing   computation methods of dS correlators.

In the beginning, let's highlight the main difference between these two cases. The integrals appearing in flat amplitude take the form
\begin{align}
\int_\mathcal{C} \prod_i P_i(\bm{z})^{\alpha_i} \nd \bm{z},\label{eq:polyintegrand}
\end{align}
where $P_i$ is polynomial and the integration region $\mathcal{C}$ is  properly chosen. 
For example, the conventional form of Feynman integral take this form, with $P_i=(l_i+q_i)^2-m_i^2$, as well as their corresponding Feynman parametrization or Baikov representation \cite{Baikov:1996iu}. On the contrary, the integrals appearing in dS correlator takes the form
\begin{align}
\int_\mathcal{C} \prod_i P_i(\bm{z})^{\alpha_i} \prod_j F_j(\bm{z})^{\beta_j} \nd \bm{z}  \, , \label{eq:fintegrand}
\end{align}
where $F_i$ are functions related to Hankel functions and step functions \cite{Chen:2017ryl}.  This poses substantial challenges in the computation of such dS correlators. Successful experience for higher precision computations of amplitudes in flat space in the past decade leads us to the Integration-By-Parts (IBP) reduction \cite{Chetyrkin:1981qh} and (first-order) Differential Equations (DE) \cite{Kotikov:1990kg,Kotikov:1991pm,Gehrmann:1999as,Bern:1993kr}  obtained by IBP. These are what we want to introduce to dS correlators. The IBP method establishes linear relations among different integrals within a integral family by integrating total differential with zero boundary. For example, for the integral family
\bea
G(a) &\equiv \int \nd^d l  \frac{ 1 }{(l^2-m^2)^a} \eea
the IBP relation gives
\bea 
0=\int  \nd^d l ~ \frac{\partial}{\partial l}\cdot l \frac{ 1 }{(l^2-m^2)^a}  ~~~&\to ~~~ (d-2a)G(a)-2 a m^2 G(a+1)=0 \, . \label{eq:IBP} 
\eea
With such relations, one can reduce any integral in the integral family into a linear combination of a selected finite set of integrals (called Master Integrals (MI)), which are the only part need to be really calculated. If the partial derivatives of the master integrals still belong to the same integral family, reducing them back to the MI yields a system of first-order differential equations (DE) of the MI.  With the boundary condition, one could solve them analytically or numerically. Moreover, when people encounter the same integral family in another scattering process, redundant computations could be avoid. The introduction of the canonical differential equation (CDE) method in \cite{Henn:2013pwa} has significantly facilitated the advancement of high-loop analytical calculations, whose DE are $\nd \log$-form. 
Then, numerical DE methods  gained attention and generalized power series expansions  \cite{Moriello:2019yhu,Bonciani:2019jyb,Frellesvig:2019byn} 
 and  Auxiliary Mass Flow (AMFlow) were developed \cite{Liu:2022mfb,Liu:2020kpc,Liu:2022tji,Liu:2017jxz} . Meanwhile, automation packages for IBP \cite{Smirnov:2019qkx,vonManteuffel:2012np,Lee:2012cn,Klappert:2020nbg,Peraro:2019svx,Wu:2023upw}  and numerical DE \cite{Liu:2022chg,Hidding:2020ytt,Armadillo:2022ugh}  have been continuously refined and widely applied  in recent years and provided immense convenience. 
Due to all these developments, IBP and DE have become prioritized tools especially for computing the most complicated amplitudes.  For instance, the recent result of a two-loop five-point process  \cite{FebresCordero:2023gjh} is obtained using IBP, numerical DE and the related packages. IBP and DE are used to be applied on the  integrals take the expression \eqref{eq:polyintegrand}. This also implies that if same forms of integrands are identified in other physical scenarios, the computational techniques developed for flat spacetime Feynman integrals may be directly applied to those situations. For instance, in dS correlator, calculations are comparatively more tractable  in the case of conformally coupled scalar due to their integrals also take the form \eqref{eq:polyintegrand}, and IBP and DE have been studied \cite{De:2023xue,Arkani-Hamed:2023kig} in this case  recently.

In this paper, the power and significant success of IBP and DE in flat spacetime  have motivated us to promote their application to dS integrals with the form \eqref{eq:fintegrand}. This extension aims to develop a systematic, efficient, and automatic technique for computing general (not only massless or conformally coupled) dS correlators. It provides also a new perspective and tool to analyze the structure and properties of dS correlator.

The structure of the paper is following. In section \ref{sec:2}, we review the Feynman rule of the dS correlator, then construct IBP of this case. In section \ref{sec:3.1}, we indicate that  each $\nd \tau_i$'s IBP are factorized, which leads to great simplification of the computation. For the so-called vertex integral family (with respect to integration of each $\tau_i$), we have constructed a smart set of bases, which will lead to further significant simplifications in the iterative reduction and DE of this integral family. In section \ref{sec:3.2}, we give a pedagogical example of iteratively reducing 1-fold Hankel function vertex integral family, to sketch the features of general cases. In section \ref{sec:3.3}, we find it coincidentally that the DE of the MI we construct automatically are $\nd \log$-form.  Subsequently, we systematically present and prove the universal formula of iterative reduction for arbitrary n-fold vertex integral families in section \ref{sec:3.4} and its $\nd \log$-form DE in section \ref{sec:3.5}. In section \ref{sec:3.6}, we quickly cover the simplest massless case. In section \ref{sec:3.7}, we give discussion of IBP and DE when they involving remaining terms come form two types of propagator $G_{\pm\pm}$ and belong to the so-called sub-sector. Finally, a brief summary and discussions are presented in the section \ref{sec:4}.

\section{Generalized IBP relations of dS correlators} \label{sec:2}

In this section, we  review the Feynman rules of in-in formalism dS correlators in cosmology and inflation. We will find  dS correlators take the form of \eqref{eq:fintegrand} and determine what terms and functions could appear as $P_i$ and $F_j$. Then we will generalize the IBP method to this case. For simplifying the discussion without losing key features, we focus on scalar field theory.

Following  the standard notation in the inflation physics, we choose the  metric with the so called conformal time $\tau$
\begin{align}
  \nd s^2=a^2(\tau)\big(-\nd \tau^2+ \nd \bm{x}^2 \big), \label{metricCF}
\end{align}
where the scale factor is chosen to be $a(\tau) = 1/(-H\tau)$ with $H$ the Hubble parameter.  Feynman rules of dS correlators (see \cite{Chen:2017ryl} for review) are similar to the case of flat space, i.e., with three basic building blocks: external wave functions, propagators and vertexes. For simplicity, let us take scalar field as an example. The mode expansion of field operator is
\begin{align}
  \varphi_a(\tau,\bm{x})=\int \frac{\nd^3\bm{k}}{(2\pi)^3}\Big[u_a(\tau,\bm{k})b_a(\bm{k})+u_a^*(\tau,-\bm{k})b_a^\dag(-\bm{k})\Big]e^{i\bm{k}\cdot\bm{x}}, \label{fieldop}
\end{align}
where the wave function $u_a(\tau,\bm{k})$ satisfies the equation of motion
\begin{align}
  u_a''(\tau,\bm{k})-\frac{2}{\tau}u_a'(\tau,\bm{k})+\Big(\bm{k}^2+\frac{m_a^2}{H^2\tau^2}\Big)u_a(\tau,\bm{k})=0\label{eq:eom}
\end{align}
with $m_a$ the mass of the field. The solution is given by 
\begin{align}
  u(\tau;k)=-i\frac{\sqrt\pi}{2}e^{i\pi(\nu/2+1/4)}H^{(d-1)/2}(-\tau)^{d/2}\text{H}_{\nu}^{(1)}(-k\tau).
\end{align}
where $\text{H}_{\nu}^{(1)}(-k\tau)$ is the Hankel function and other parameters are 
$k=|\bm{k}|$, $v=\sqrt{\frac{d^2}{4}-\frac{m^2}{H^2}}$ and $d=3$.
The Hankel functions satisfy\footnote{The definition of $\text{H}_{\nu}^{(1,2)}$ in textbook is 
$\text{H}_{\nu}^{(1)}(z)= \text{J}_{\nu}(z)+ i \text{Y}_{\nu}(z)$ and $\text{H}_{\nu}^{(2)}(z)= \text{J}_{\nu}(z)- i \text{Y}_{\nu}(z)$. When $z,\nu$ are real number, it is obviously that $\text{H}_{\nu}^{(2)}(z)=(\text{H}_{\nu}^{(1)}(z))^*$. The second line  of \eqref{eq:hankel} is a generalization of above result when $z,\nu$ are complex number.} 
\begin{align}
  & \partial_\tau^2 \text{H}_{\nu}^{(1,2)}(-k\tau)+\frac{1}{\tau} \partial_\tau\text{H}_{\nu}^{(1,2)}(-k\tau)+\Big(k^2-\frac{\nu^2}{H^2 \tau^2}\Big)\text{H}_{\nu}^{(1,2)}(-k\tau)=0 \, ,\nn\\
  & \text{H}_{\nu}^{(1)}(-k\tau)= \left(\text{H}_{\nu^\star}^{(2)}(-k^\star\tau^\star) \right)^\star \, ,\nn\\
  &\tau\to 0_+ : \text{H}_{\nu}^{(1)}(-k\tau) \sim  - \frac{i 2^\nu \Gamma [\nu]}{\pi}  (-k \tau)^{-\nu}  \,, \ \text{for Re}[\nu] >0 \nn\\
  &\tau\to + \infty : \text{H}_{\nu}^{(1)}(-k\tau) \sim \sqrt{\frac{2}{\pi }} (-k \tau)^{ -\frac{1}{2}}  e^{-i k \tau-i \pi  \left(\nu/2+1/4\right)}
  \label{eq:hankel}
\end{align}

Using \eqref{fieldop} we can get various propagators. The bulk-to-bulk propagators are given by
\begin{align}
G_>(k;\tau_1,\tau_2)\equiv&~u(\tau_1,k)u^*(\tau_2,k),\nn\\
G_<(k;\tau_1,\tau_2)\equiv&~u^{*}(\tau_1,k)u(\tau_2,k).\\
G_{++}(k;\tau_1,\tau_2)=&~G_>(k;\tau_1,\tau_2)\theta(\tau_1-\tau_2)+G_<(k;\tau_1,\tau_2)\theta(\tau_2-\tau_1),\nn\\
G_{+-}(k;\tau_1,\tau_2)=&~G_<(k;\tau_1,\tau_2),\nn\\
G_{-+}(k;\tau_1,\tau_2)=&~G_>(k;\tau_1,\tau_2),\nn\\
G_{--}(k;\tau_1,\tau_2)=&~G_<(k;\tau_1,\tau_2)\theta(\tau_1-\tau_2)+G_>(k;\tau_1,\tau_2)\theta(\tau_2-\tau_1),  \label{eq:proptype}
\end{align}
The bulk-to-boundary propagators are non-vanish only when the field is massless 
\begin{align}
G_{+}(k;\tau)\equiv G_{+\pm}(k;\tau_1,0)=\frac{H^2}{2k^3}(1-ik\tau)e^{+ik\tau} ,\nn\\
G_{-}(k;\tau)\equiv G_{-\pm}(k;\tau_1,0)=\frac{H^2}{2k^3}(1+ik\tau)e^{-ik\tau} .
\end{align}
It is necessary to notice that when people consider the asymptotic past $\tau\to-\infty$, i$\epsilon$-prescription \cite{Chen:2017ryl} should be taken into consideration for good behavior. Practically, it is equivalent to attach an external factor $e^{\epsilon k \tau}$ to $G_>$ and $G_<$  making propagators to be exponentially suppressed to zero when $\tau\to-\infty$.
For Feynman rules of  vertices, some examples are listed as  follows:
\begin{align}
-\frac{\lambda}{24}a^4(\tau)\varphi^4  \to & \mp i\lambda \int_{-\infty}^{0}\nd\tau\,a^4(\tau)\cdots \nn\\
-\frac{\lambda}{6}a^2(\tau)\varphi(\partial_i\varphi)(\partial_i\varphi)\to &
\pm \frac{i\lambda}{3} (k_{12}+k_{23}+k_{13})\int_{-\infty}^{0}\nd\tau\,a^2(\tau)\cdots,~~~~k_{ij}\equiv \bm{k}_i\cdot \bm{k}_j \nn\\
-\frac{\lambda}{6}a^2(\tau)\varphi\varphi'^2 \to & \mp \frac{i \lambda}{3}\int_{-\infty}^{0}\nd\tau\,a^2(\tau)G_{+a_1}(k_1;\tau,\tau_1)\prod_{i=2,3}\big[\partial_{\tau}G_{+a_i}(k_i;\tau,\tau_i)\big]\nn\\
      &+\text{2 permutations},
\end{align}

From above discussions, one can see that dS correlators are indeed the form of
\eqref{eq:fintegrand} with following correspondence: 
\begin{align}
\int_\mathcal{C} \nd \bm{z}&= \int_{-\infty}^0 \nd \tau_i\int_{-\infty}^{\infty} \nd \bm{l}_i \, , 
~~~~\text{$\bm{l}_i$ for loop momentum} \, , 
\nn\\
P_i & = \tau_j, ~~ \text{polynomial of loop momentum} \, , \nn\\
F_i & = e^{i k \tau}, ~~ \text{H}_{\nu}^{(1,2)} (-k \tau), ~~ \partial_{\tau} \text{H}_{\nu}^{(1,2)}(-k \tau), ~~ \theta(\tau_j-\tau_k) \, . \label{eq:dSintegrand}
\end{align}
We call a integral family defined by \eqref{eq:dSintegrand} a \textbf{dS integral family}.

To apply IBP method, we need to check the  dS integral family satisfying  the proper boundary conditions. 
For the loop integration, just like the flat space, we will  use the dimension regularization,
thus the IBP relations involving $\bm{l}_i$ satisfy automatically. The new feature is the conformal time integration. First with the $i\epsilon$-prescription, the integrands equal zero at the boundary $\tau_i=-\infty$. For the boundary $\tau_i=0$, one could add a regulator $\tau_i^{\delta_i}$ if it is divergence, then it is again zero. The regulator for the divergent case is necessary for the IBP method to apply,
since only with it the integration is well defined. Physical information is coded at the series expansion of $\delta$ (just like dimension regulator $\epsilon$ for flat space), for example, $1/\delta$ term in the expansion. In IBP-based differential equations method for flat spacetime, obtaining result in the expansion of regulator is also a usual practice. Since $a(\tau)\sim \frac{1}{ \tau}$, power function of  $\tau$ automatically appears in the 
integrand. 
With these considerations, we conclude that the IBP method can be applied to dS integral family. 
Let us do one simple  example 
\begin{align}
0=\int_\mathcal{C} \nd \left(\tau^{\nu_0} \text{H}_{\nu}^{(1)} (-k \tau) \right), \label{eq:ibpeg1}
\end{align}
gives
\begin{align}
0=\nu_0 \int_\mathcal{C} \tau^{\nu_0-1} \text{H}_{\nu}^{(1)} (-k \tau)\nd \tau + \int_\mathcal{C} \tau^{\nu_0} \partial_\tau\text{H}_{\nu}^{(1)} (-k \tau)\nd \tau    , \label{eq:ibpeg2}
\end{align}
 One may notice that applying the derivative operator to $\partial_{\tau} \text{H}_{\nu}^{(1,2)}(-k \tau)$ will give $\partial_{\tau}^2 \text{H}_{\nu}^{(1,2)}(-k \tau)$, which is not in the dS integral family \eqref{eq:dSintegrand}. However, using the relation \eqref{eq:hankel} we can express $\partial_{\tau}^2 \text{H}_{\nu}^{(1,2)}(-k \tau)$ as the linear combination of $ \text{H}_{\nu}^{(1,2)}(-k \tau)$ and 
 $\partial_\tau \text{H}_{\nu}^{(1,2)}(-k \tau)$. For $\partial_{k_i}$, similar things also happen. For $\theta$-function, the derivative leads to $\delta$-function which can be easily integrated out. Overall, IBP relations 
 obtained from a defined dS integrals family will be a closed set of equations, which can be solved 
 systematically.


\section{IBP of $\nd\tau_i$: vertex integral family}\label{sec:3}

The integral family of $\nd \tau_i$ plays a key role in IBP of dS integrals and we will focus on its special properties in this section. 


\subsection{vertex integral family and factorization of $\nd \tau_i$ IBP} \label{sec:3.1}

Before going to details, let us notice that from \eqref{eq:dSintegrand}  for most terms, (at most times) different $\tau_i$'s are factorized. There is only one exception, i.e., the $\theta(\tau_i-\tau_j)$, which comes from $G_{++}$ and $G_{--}$. Thus we  define a \textbf{vertex integral family} corresponding to these good integrands without $\theta$-function: 
\begin{align}
&V(\nu_0,a_1,a_2,\cdots,a_n)= V(\nu_0,\bm{a})=\int  \hat{V}(\nu_0,\bm{a};\tau) \nd\tau \nn\\
&~~~~\equiv\int_{-\infty}^0 \tau^{\nu_0} e^{ik_0 \tau} 
 \prod_i h(\nu_i,a_i;-k_i\tau)\nd \tau ,~~~~~ h\equiv h^{(1,2)} \, , \ a_i=0,1 \, , \nn\\
 &h^{(1\text{ or }2)}(\nu,0;-k\tau)\equiv  (-k\tau)^{-\nu} \text{H}_{\nu}^{(1)}(-k \tau) ~ (\text{or }\text{H}_{\nu^\star}^{(2)} )\propto \tau^{-\frac{3}{2}-\nu} u ~ (\text{or } u^*),\nn\\
& h^{(1,2)}(\nu,1;-k\tau)\equiv - \frac{1}{k} \partial_\tau h^{(1,2)}(\nu,0;-k\tau)  \, .
\label{eq:vtxfamily}
\end{align}
In the definition we have omitted the dependence on  $k_0,k_i,\nu_i$ for simplicity. 
We introduce also a factors $(-k\tau)^{-\nu}$ in front of Hankel function to the definition of $h$. This factor is one of the most important construction in this paper.  One can check that   the differential equation of $h$ 
\begin{align}
  h''(\nu,0;-k\tau)+\frac{1}{\tau}(2\nu+1)h'(\nu,0;-k\tau)+k^2 h(\nu,0;-k\tau)=0, \label{eq:eom1}
\end{align}
can be rewritten as
\begin{align}
 \partial_{\tilde{\tau}}^2 h(\nu,0;\tilde{\tau})+\frac{1}{\tilde{\tau}}(2\nu+1)\partial_{\tilde{\tau}}h(\nu,0;\tilde{\tau})+ h(\nu,0;\tilde{\tau})=0, ~~~\tilde{\tau}=-k\tau\label{eq:eom2}
\end{align}
and $ h(\nu,1;-k\tau)$ is defined by $\partial_{\tilde{\tau}}h(\nu,0;-k\tau)$. To cancel the $1/\tau^2$  term in the differential equation \eqref{eq:hankel}, one need multiply  a prefactor  $(-k\tau)^{\pm\nu}$ in the defination of h. Here we choose $(-k\tau)^{-\nu}$.  Another key point of \eqref{eq:eom2} is that $h$ depends only on the combination of $k\tau$, not individual $k$ and $\tau$.  As it will be seen, this construction  extremely simplify the IBP relations of vertex integral family. We even find  coincidentally that this definition  automatically gives $\nd \log$-form DE of all vertex integral family. We will show the details later. 

Due to the fact that u and $u^\star$, or equivalently, $\text{H}_{\nu}^{(1)}(-k \tau)$ and $\text{H}_{\nu^\star}^{(2)}(-k \tau)$ satisfy the same differential equation for the natural condition $k\in \mathbb{R}$ and $\nu=\pm\nu^\star$ (or says $\nu$ are real or pure imaginary)\footnote{One can see it easily by checking the first line  of \eqref{eq:hankel} where only $\nu^2$ appears in the differential equation. With this condition in the definition of $h(\nu,0;-k\tau)$, the prefactor $(-k\tau)^{-\nu}$ will be same for both $\text{H}_{\nu}^{(1)}(-k \tau)$ and $\text{H}_{\nu^\star}^{(2)} $. Otherwise, the prefactor needs to be changed accordingly. }, they also share the same properties under IBP. Thus we use the same symbol to denote both of them here for convenience. 
We call the integral family consists of these $V(\nu_0+a_0,a_1,a_2,\cdots,a_n)$s for selected $\nu_i$ and $k_i$ an \textbf{n-fold} (Hankel function) vertex integral family.

To apply IBP method, we need know the action of differential operators on  $h(\nu,a,-k\tau)$. For $\partial_\tau$, we have
\begin{align}
 \partial_\tau h(\nu,0,-k\tau) &= -k h(\nu,1,-k\tau) \, ,\nn\\
 \partial_\tau h(\nu,1,-k\tau) &= -k \Big[ \frac{1}{k\tau}(2\nu+1)h(\nu,1;-k\tau) - h(\nu,0;-k\tau) \Big] \, .\label{eq:ibpdt}
\end{align}
To construct IBP relation at loop-level corresponding to $\nd l_i$, and also construct DE with respect to $k_i$, we also need know the result of applying $\partial_{k_i}$ to $h(\nu,a,-k\tau)$. One can use the properties of Hankel function to get the result, but a much simpler way is to notice that $h(\nu,a;\tilde{\tau})$'s dependence on $k$ and $\tau$ are only mediated through $\tilde{\tau}$ as given in 
\eqref{eq:eom2}. Thus one immediately get
\begin{align}
 \partial_k h(\nu,0,-k\tau) &= -\tau h(\nu,1,-k\tau) \, ,\nn\\
 \partial_k h(\nu,1,-k\tau) &= -\tau \Big[ \frac{1}{k\tau}(2\nu+1)h(\nu,1;-k\tau) - h(\nu,0;-k\tau) \Big] \, . \label{eq:ibpdk}
\end{align}
With \eqref{eq:ibpdt} and \eqref{eq:ibpdk}, IBP relations and DE corresponding to $\partial_k$ could be easily constructed. 

Now we want to show some important properties of this IBP system. Let us consider expressions involving only  $G_{+-}$ or/and $G_{-+}$ type propagators. In this case, the correlator factorizes to 
\begin{align}
\int \prod_j \left[\hat{V}_j(\cdots;\tau_j) \nd \tau_j\right]  f(l_1,\cdots,l_L) \prod_i\nd l_i
\end{align}
where $f$ is polynomial of $l_i$ and $\hat{V}$ is given in \eqref{eq:vtxfamily}.
The IBP relation with respect to a selected $\tau_k$ is given by
\begin{align}
\int \left(\partial_{\tau_k} \hat{V}_k(\cdots;\tau_k) \nd \tau_k\right)  \times \prod_{j\neq k} \left(\hat{V}_j(\cdots;\tau_j) \nd \tau_j\right)  f(l_1,\cdots,l_L;k_1\cdots,k_E) \prod_i\nd l_i
\end{align}
which does not affect other parts. We say the IBP relations with respect to different $\tau_i$ are factorized.  

Now let's consider $G_{++}$ and $G_{--}$ type propagators. Its factorization is not as good as $G_{\pm \mp}$, since derivative acting on $\theta$-function leads to  the term
\begin{align}
\int  \hat{V}_i(\cdots;\tau_i) \delta(\tau_i-\tau_j) \hat{V}_j(\cdots;\tau_j)  \nd \tau_i\nd\tau_j\times \cdots = \int  \hat{V}_i(\cdots;\tau_i)\hat{V}_j(\cdots;\tau_i) \nd\tau_i\times\cdots, \label{eq:dtheta0}
\end{align}
where naively $(n_i+n_j)$-fold Hankel function emerges. However, as we will show later, in fact,
it becomes $(n_i+n_j-2)$-fold Hankel function or vanishes completely. Fortunately, the new term is belong a sub-sector with less vertex integral family, by which we means the IBP of the sub-sector will not invovle the orignal higher sector. So we can still reduce each vertex integral family individually. And these remaining terms belong to sub-sector are regarded as known results in this part of reduce which have been reduced using the IBP-reduction of sub-sector.

Now we consider the merging happened in \eqref{eq:dtheta0} by considering the symmetry in  $G_{\pm \pm}$ more carefully.
There are four cases could appear in the IBP system which involves massive $G_{\pm \pm}$:
\begin{align}
&\left[h^{(a)}(\nu,0,-k\tau_i) (\partial_{\tau_i}\theta_{ij}) h^{(3-a)}(\nu,0,-k\tau_j) + h^{(3-a)}(\nu,0,-k\tau_i) (\partial_{\tau_i}\theta_{ji}) h^{(a)}(\nu,0,-k\tau_j) \right] \times \cdots  \nn\\
&\left[h^{(a)}(\nu,1,-k\tau_i) (\partial_{\tau_i}\theta_{ij}) h^{(3-a)}(\nu,0,-k\tau_j) + h^{(3-a)}(\nu,1,-k\tau_i) (\partial_{\tau_i}\theta_{ji}) h^{(a)}(\nu,0,-k\tau_j) \right] \times \cdots  \nn\\
&\left[h^{(a)}(\nu,0,-k\tau_i) (\partial_{\tau_i}\theta_{ij}) h^{(3-a)}(\nu,1,-k\tau_j) + h^{(3-a)}(\nu,0,-k\tau_i) (\partial_{\tau_i}\theta_{ji}) h^{(a)}(\nu,1,-k\tau_j) \right] \times \cdots  \nn\\
&\left[h^{(a)}(\nu,1,-k\tau_i) (\partial_{\tau_i}\theta_{ij}) h^{(3-a)}(\nu,1,-k\tau_j) + h^{(3-a)}(\nu,1,-k\tau_i) (\partial_{\tau_i}\theta_{ji}) h^{(a)}(\nu,1,-k\tau_j) \right] \times \cdots  \nn\\
& ~~~ \theta_{ij} \equiv \theta(\tau_i-\tau_j)  \,, \ \ a=1,2 \, .
\end{align}
The integration $\int \nd\tau_i \nd \tau_j$ of first and fourth cases will vanish due to
\begin{align}
&\int \nd \tau_i \nd \tau_j \left[h^{(a)}(\nu,0,-k\tau_i) h^{(3-a)}(\nu,0,-k\tau_j) - h^{(3-a)}(\nu,0,-k\tau_i) h^{(a)}(\nu,0,-k\tau_j) \right]  \delta(\tau_i-\tau_j) \times \cdots  \nn\\
&= \int \nd \tau_i \left[h^{(a)}(\nu,0,-k\tau_i) h^{(3-a)}(\nu,0,-k\tau_i) - h^{(3-a)}(\nu,0,-k\tau_i) h^{(a)}(\nu,0,-k\tau_i) \right] \times \cdots =0 \nn\\
&\int \nd \tau_i \nd \tau_j \left[h^{(a)}(\nu,1,-k\tau_i) h^{(3-a)}(\nu,1,-k\tau_j) - h^{(3-a)}(\nu,1,-k\tau_i) h^{(a)}(\nu,1,-k\tau_j) \right]  \delta(\tau_i-\tau_j) \times \cdots  \nn\\
&= \int \nd \tau_i \left[h^{(a)}(\nu,1,-k\tau_i) h^{(3-a)}(\nu,1,-k\tau_i) - h^{(3-a)}(\nu,1,-k\tau_i) h^{(a)}(\nu,1,-k\tau_i) \right] \times \cdots =0  \, . \label{eq:vanishR}
\end{align}
The integration $\int \nd\tau_i \nd \tau_j$ of second and third cases give
\begin{align}
&  - (-1)^a\int \nd \tau_i \left[F(-k\tau_i) \right]  \times \cdots =  - (-1)^a\int \nd\tau_i ~e^{ \pi \text{Im}[\nu]} \frac{  4 i }{\pi}  (-k\tau_i)^{-2\nu-1} \times \cdots \nn\\
&  + (-1)^a \int \nd \tau_i \left[F(-k\tau_i) \right]  \times \cdots =  + (-1)^a \int \nd\tau_i ~e^{ \pi \text{Im}[\nu]}\frac{  4 i }{\pi}  (-k\tau_i)^{-2\nu-1} \times \cdots \nn\\
& F(-k\tau_i) =h^{(1)}(\nu,1,-k\tau_i)  h^{(2)}(\nu,0,-k\tau_i) - h^{(2)}(\nu,1,-k\tau_i)  h^{(1)}(\nu,0,-k\tau_i) \label{eq:pinchR} 
\end{align}
for real or imaginary $\nu$, where we have used the property of Hankel function
\begin{align}
& F(-k\tau_i) = e^{ \pi \text{Im}[\nu]} \frac{  4 i }{\pi}  (-k\tau_i)^{-2\nu-1} \,  .
\end{align}
One can also check it that the derivative of $F(-k\tau_i)$ gives $\partial_{- k \tau} F(-k\tau) = - \frac{2\nu+1}{- k \tau} F(- k \tau)$, then, $F(-k\tau_i) = C (-k\tau_i)^{-2\nu-1}$ and $C$ can be determined by asymptotic behavior of Hankel functions on the boundary. Then, two vertex are "pinched" together and give  a $(n_i+n_j-2)$-fold vertex integral family, not $(n_i+n_j)$-fold from the naive observation in \eqref{eq:dtheta0}.

For massless $G_{\pm \pm}$, if one choose to keep the combination of the two terms in these propagator, there are also "vanishing" case and "pinched” case. "Vanishing” case comes from
\begin{align}
 \int \nd \tau_i \nd \tau_j \left[ e^{ik \tau_i} (\partial_{\tau_i} \theta_{ij}) e^{ - ik \tau_j} +  e^{-ik \tau_i} (\partial_{\tau_i} \theta_{ji}) e^{  ik \tau_j} \right]\times \cdots = 0
\end{align}
and "pinched" cases comes from
\begin{align}
&\int \nd \tau_i \nd \tau_j \left[  (\partial_{\tau_i}e^{ik \tau_i}) (\partial_{\tau_i} \theta_{ij}) e^{ - ik \tau_j} 
+  (\partial_{\tau_i}e^{-ik \tau_i}) (\partial_{\tau_i} \theta_{ji}) e^{  ik \tau_j} \right] \times \cdots = \int \nd \tau_i 2 i k \times \cdots \nn\\
&\int \nd \tau_i \nd \tau_j \left[  e^{ik \tau_i} (\partial_{\tau_i} \theta_{ij})\partial_{\tau_i} e^{ - ik \tau_j} 
+  e^{-ik \tau_i} (\partial_{\tau_i} \theta_{ji}) \partial_{\tau_i}e^{  ik \tau_j} \right] \times \cdots = -\int \nd \tau_i 2 i k \times \cdots 
\end{align}
One can also choose to handle the two terms in massless propagator individually.

With the above discussion,  we can reduce each vertex integral family individually: 
\begin{align}
\int \prod_i \nd\tau_i \hat{V}_i    \times \cdots =  \left[ \int \prod_i \Big(\nd\tau_i \sum_j c^{(i)}_j \hat{f}^{(i)}_j  \Big)   \times \cdots \right] + R, \label{eq:factorize}
\end{align}
where $\hat{f}^{(i)}_j$ is the  integrand of the MI of the integral family $V_i$ belong to,   $"\cdots"$ part could contain $\theta(\tau_i-\tau_j)$, and we use $R$ to denote remaining terms. $R$ terms come from pinching two vertex connected via $G_{\pm\pm}$ together. Since the reduction of these $R$ terms will not involve the family before pinching,  it belongs to a sub-sector and can treat as known in the reduction of family $V_i$ in \eqref{eq:factorize}. This form leads  following simplifications in the computation:
\begin{itemize}
    \item \textbf{Factorization of IBP relations of vertex}: 
    We say that the IBP relations of different $\nd \tau_i$ are factorized in the $(t,\bm{k})$ space of dS integral, since we can perform the IBP reduction individually for each of them, as shown in \eqref{eq:factorize}. Since all vertex integral family take the similar expression \eqref{eq:vtxfamily}, the reduction result can be used repeatedly for different vertex, correlator and theory. For example, once people get the IBP reduction of a n-fold vertex integral family, it can be directly applied to all n-fold vertex appears in other place (if without external symmetry).
\begin{figure}[h]
\centering
   \quad  \includegraphics[width=5.5 in]{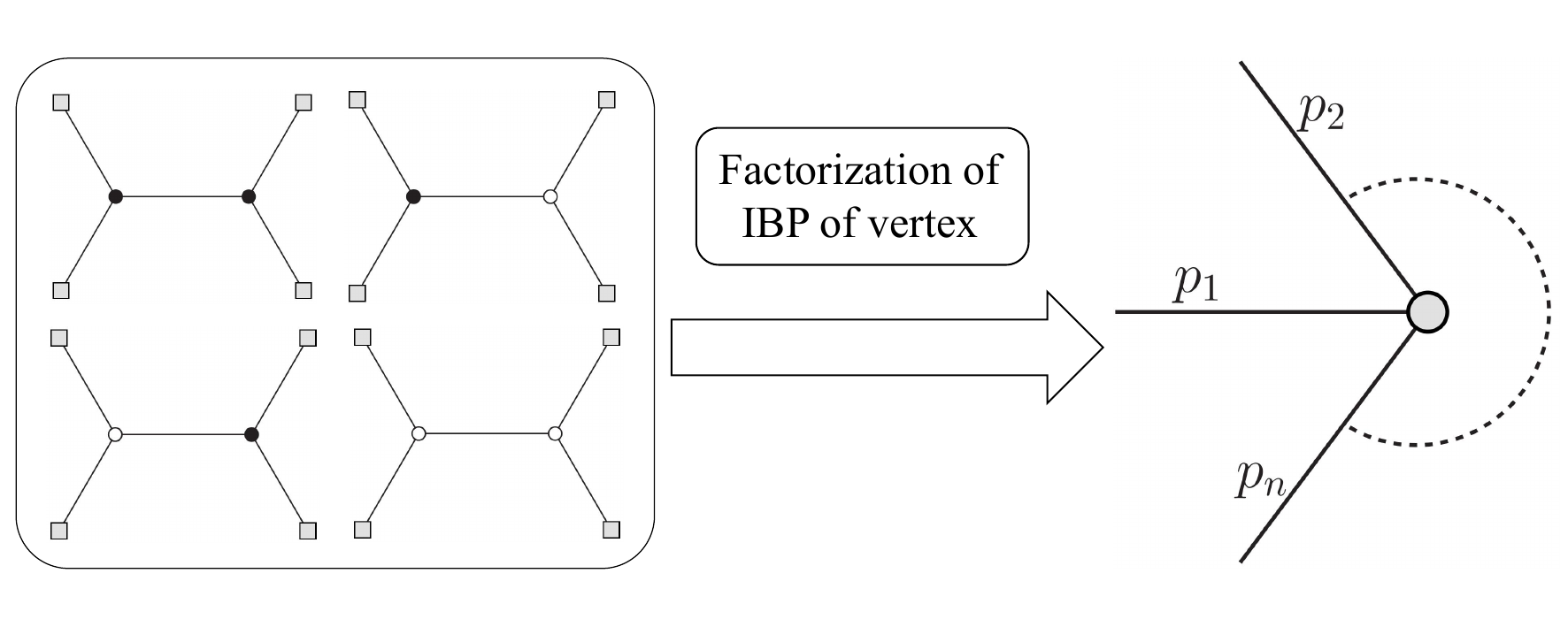}
    \caption{Factorization of IBP relations of vertex: all reductions with respect to $\nd \tau$ can be achieved by reducing individual families of vertex integrals separately.}
    \label{fig:factorization}
\end{figure}
    \item At loop-level, one need to further consider the IBP relations of $\nd l_i$. Notice that terms outside of $\hat{V}_i$ and $\theta$-function are independent of $\tau_i$. Due to this, 
    one can reduce the family of all $\nd \tau$ integrals to MI first, to get $\nd\tau_i$-reduced IBP relations. Then, solving these $\nd\tau_i$-reduced IBP relations of $\nd l_i$ will give the complete result of reduction.
 \begin{figure}[h]
\centering
   \quad  \includegraphics[width=5.5 in]{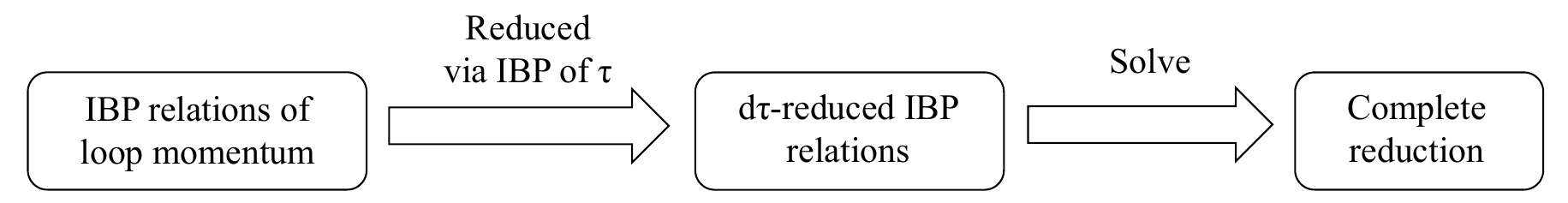}
    \caption{Steps of loop-level reduction}
    \label{fig:factorization}
\end{figure}
        In this progress, differential equations of $\nd \tau$ integrals family could serve for directly giving $\nd\tau_i$-reduced IBP relations. For example, consider total derivative with respect to $l$ for the following loop integrals
\begin{align}
\int \nd \left(\frac{1}{|l||l-p_1|\cdots} \bm{f} \right) =0  \, ,\ \  \bm{f}=\{f_1,f_2,\cdots\}
\end{align}
where the $\bm{f}$ are  MI of the integrals family of all $\nd \tau_i$. Then, we have a series of $\nd\tau_i$-reduced IBP relations
\begin{align}
\int \left(\nd  \frac{1}{|l||l-p_1|\cdots}\right) \bm{f}+ \frac{1}{|l||l-p_1|\cdots} \left( \Omega.\bm{f} \nd l \right) = 0 \,, \ \ \nd \bm{f} = \Omega.\bm{f}\nd l \label{eq:egdtRibp}
\end{align}
where the $ \Omega$ is the differential equation matrix of $\bm{f}$ with respect to $l$. \eqref{eq:egdtRibp} is directly the so called $\nd \tau$-reduced IBP relation, because the $\nd \tau$ integration part of each integral in these relations are kept as master integrals.
\end{itemize}

In the rest of this section, we will show some examples for n-fold vertex integral family and sketch more features of this IBP system. Following the idea similar to iterative reduction \cite{Chen:2022jux,Chen:2022lue} in flat space, we will  further see that the infinite number of integrals in a vertex integral family can be reduced iteratively to MI once we have solved the iterative relations from a finite linear system. Furthermore, we get the universal expression of these iterative reductions for any n-fold vertex integral family, as well as $\nd \log$-form DE of MI. We also will give formulas and discussion for reduction with remaining terms come from $G_{\pm\pm}$.

\subsection{1-fold vertex integral family: iterative reduction} \label{sec:3.2}
Consider 1-fold vertex integral family
\begin{align}
V(\nu_0,a_1)&=\int_{-\infty}^0 \tau^{\nu_0} e^{ik_0 \tau} 
  h(\nu_1,a_1;-k_1\tau)\nd \tau ,~~~~a_i=0,1 \, , 
\end{align}
From now on we set $H$ to be $1$ for convenience. We use $eq[\nu_0,\bm{a}]$ to denote the IBP relation
\begin{align}
\int\nd \hat{V}(\nu_0,\bm{a};\tau)=0,
\end{align}
then we have IBP relations
\begin{align}
eq[\nu_0,0]&: ~ i k_0 V\left(\nu _0,0\right)+\nu _0 V\left(\nu _0-1,0\right)-k_1 V\left(\nu _0,1\right)=0   \nn\\
eq[\nu_0,1]&:~ i k_0 V\left(\nu _0,1\right)+\nu _0 V\left(\nu _0-1,1\right) + k_1 V\left(\nu _0,0\right)+\left(-2 \nu _1-1\right) V\left(\nu_0-1,1\right)=0.
\end{align}
We have numerically evaluated these integrals at their convergent region and verified these two IBP relations to be right.
Defining
\begin{align}
\bm{f}^{(a_0)}=\{V\left(\nu_0+a_0,0\right),V\left(\nu_0+a_0,1\right)\},
\end{align}
the IBP relation corresponding to a selected $\nu_0$ can be expressed in the form of matrix as
\begin{align}
&\left(M_{1}^{(1)}+\nu_0 \mathbb{I}_{2} \right).\bm{f}^{(-1)}+\left(M_{0}^{(1)}+ik_0\mathbb{I}_{2}\right).\bm{f}^{(0)} =0  \nn\\
&M_{1}^{(j)}=-\frac{2\nu_j+1}{2}\left(\mathbb{I}_{2}-\sigma_3 \right)=\left(
\begin{array}{cc}
 0 & 0 \\
 0 & -2\nu_j-1 \\
\end{array}
\right) , ~~~~ M_{0}^{(j)}= - i k_j \sigma_2 =\left(
\begin{array}{cc}
 0 & -k_j \\
 k_j & 0 \\
\end{array}
\right),  \label{eq:matrixibp}
\end{align} 
where $\sigma_{1,2,3}$ are the Pauli matrices. Traditionally, taking $\nu_0$ to be $\nu_0+a_0$ with different $a_0$, we can get a series of IBP relations. Solving them give the reduction of this integral family. As the consequence, we find there are two MI in this system, which can be chosen as $\bm{f}^{(0)}$. However, here we obviously have a simpler method. People only need to solve iterative relation
\begin{align}
\bm{f}^{(-1)}=A_-(\nu_0).\bm{f}^{(0)},
\end{align}
Then, all integrals in the family can be reduced iteratively by
\begin{align}
\bm{f}^{(-n)}&=\left(\prod_{i=n-1}^0 A_-(\nu_0-i) \right) . \bm{f}^{(0)},\nn\\
\bm{f}^{(n)}&=\left(\prod_{i=n-1}^0 A_+(\nu_0+i)\right) . \bm{f}^{(0)},~~~~ A_+(\nu_0)\equiv \left(A_-(\nu_0+1)\right)^{-1},
\end{align}
For 1-fold vertex integral family, $A_\pm$ can be immediately solved and given as following
\begin{align}
&A_-(\nu_0) = \left(
\begin{array}{cc}
 -\frac{i k_0}{\nu _0} & \frac{k_1}{\nu _0} \\
 \frac{k_1}{-\nu _0+2 \nu _1+1} & -\frac{i k_0}{\nu _0-2 \nu _1-1} \\
\end{array}
\right) \nn\\
&A_+(\nu_0)=\left(
\begin{array}{cc}
 \frac{i k_0 \left(\nu _0+1\right)}{k_0^2-k_1^2} & \frac{k_1 \left(\nu _0-2 \nu _1\right)}{k_0^2-k_1^2} \\
 \frac{k_1 \left(\nu _0+1\right)}{k_1^2-k_0^2} & \frac{i k_0 \left(\nu _0-2 \nu _1\right)}{k_0^2-k_1^2} \\
\end{array}
\right) \label{eq:1Hreduce}
\end{align}
For $k_0=0$, they will be simplified to
\begin{align}
&A_-(\nu_0) =\left(
\begin{array}{cc}
 0 & \frac{k_1}{\nu _0} \\
 \frac{k_1}{-\nu _0+2 \nu _1+1} & 0 \\
\end{array}
\right) ,~~~~~A_+(\nu_0)=\left(
\begin{array}{cc}
 0 & -\frac{\nu _0-2 \nu _1}{k_1} \\
 \frac{\nu _0+1}{k_1} & 0 \\
\end{array}
\right), \label{eq:1Hreducek00}
\end{align}

\subsection{1-fold vertex integral family: $\nd\log$-form DE}  \label{sec:3.3}
Using above reduction we can establish the DE for the MI by acting 
 $\partial_{k_0}$ and $\partial_{k_1}$ on $\bm{f}^{(0)}$. They gives
\begin{align}
&\partial_{k_0}\bm{f}^{(0)}=\left(
\begin{array}{c}
 i V\left(\nu _0+1,0\right) \\
 i V\left(\nu _0+1,1\right) \\
\end{array}
\right) = i A_+(\nu_0+1)\bm{f}^{(0)} \,, \nn\\
&\partial_{k_1}\bm{f}^{(0)}=\left(
\begin{array}{c}
 -V\left(\nu _0+1,1\right) \\
 \frac{\left(-2 \nu _1-1\right) V\left(\nu _0,1\right)}{k_1}+V\left(\nu _0+1,0\right) \\
\end{array}
\right)=\left(\frac{1}{k_1}M^{(1)}_1-i  \sigma_2.A_+(\nu_0+1) \right).\bm{f}^{(0)}.
\end{align}
Evidently, the partial derivatives do not give rise to any special functions beyond the original family of functions, this implies that DE method corresponding to $\partial_{k_i}$ can be applied for such integral family.
Let's reduce the right hand side of these equations with the iterative relation we get in \eqref{eq:1Hreduce}. It gives
\begin{align}
&\partial_{k_i}\bm{f}^{(0)}= \Omega_{k_i}.\bm{f}^{(0)}, \nn\\
&\Omega_{k_0}=\left(
\begin{array}{cc}
 -\frac{k_0 \left(\nu _0+1\right)}{k_0^2-k_1^2} & 
 \frac{i k_1 \left(\nu _0-2 \nu _1\right)}{k_0^2-k_1^2} \\ -\frac{i k_1 \left(\nu _0+1\right)}{k_0^2-k_1^2} & -\frac{k_0 \left(\nu _0-2 \nu _1\right)}{k_0^2-k_1^2} \\
\end{array}
\right)  ,\nn\\
& \Omega_{k_1}=\left(
\begin{array}{cc}
 \frac{k_1 \left(\nu _0+1\right)}{k_0^2-k_1^2} &  
 -\frac{i k_0 \left(\nu _0-2 \nu _1\right)}{k_0^2-k_1^2} \\ -\frac{i k_0 k_1 \left(\nu _0+1\right)}{k_1^3-k_0^2 k_1} & \frac{k_0^2 \left(2 \nu _1+1\right)-k_1^2 \left(\nu _0+1\right)}{k_1^3-k_0^2
   k_1} \\
\end{array}
\right).
\end{align}
In flat space, canonical DE play an important role in analytical calculation of loop integral. It takes the form
\begin{align}
\nd \bm{f} = \ep (\nd\Omega).\bm{f}
\end{align}
with $\nd \Omega$ in $\nd \log$-form. It is interesting that we find in 1-fold vertex integral family, in the MI we defined, the DE automatically is in $\nd \log$-form. Although it is subtly different from canonical DE, we report this elegant result as it may offer potential assistance in understanding the mathematical structure of dS integrals in the future. The DE can be written in $\nd \log$-form as
\begin{align}
&\nd \bm{f}^{(0)}= \nd\Omega \bm{f}^{(0)} \nn\\
&\nd\Omega = \sum_{i=0,1}\tilde{\Omega}_{k_i} \nd k_i = C_1 \nd\log(k_1) + C_2 \nd\log[(k_0-k_1)(k_0+k_1)]+C_3 \nd\log\left(\frac{k_0+k_1}{k_0-k_1}\right)\nn\\
&C_1=\left(
\begin{array}{cc}
 0 & 0 \\
 0 & -2 \nu _1-1 \\
\end{array}
\right) ,    ~~   C_2=\left(
\begin{array}{cc}
 \frac{1}{2} \left(-\nu _0-1\right) & 0 \\
 0 & \frac{1}{2} \left(2 \nu _1-\nu _0\right) \\
\end{array}
\right)\nn\\
&C_3=\left(
\begin{array}{cc}
 0 & -\frac{1}{2} i \left(\nu _0-2 \nu _1\right) \\
 \frac{1}{2} i \left(\nu _0+1\right) & 0 \\
\end{array}
\right).
\end{align}
where $\nd f \equiv \sum_i  \partial_{k_i} f ~ \nd k_i$, and $\nd \nu_i$ is not included (i.e., we treat $\nu_i$ as fixed parameters).

\subsection{n-fold vertex integral family: universal formula of iterative reduction}  \label{sec:3.4}

Now we consider  n-fold vertex integral family
\begin{align}
V(\nu_0,a_1,\cdots,a_n)&=\int_{-\infty}^0 \tau^{\nu_0} e^{ik_0 \tau} 
  \prod_{i=1}^n h(\nu_i,a_i;-k_i\tau)\nd \tau ,~~~~ a_i=0,1 \,.
\end{align}
It has $2^n$ MI, which can be chosen as
\begin{align}
f^{(0)}_{\bm{a}}=V(\nu_0,\bm{a}), ~~~\bm{a}=a_1,\cdots,a_n,~~~\forall a_i=,0,1.
\end{align}
and together denoted as $\bm{f}^{(0)}$. As a vector with $2^n$ components, 
the ordering is given by 
$n$ indices $a_i$ according to  the natural binary  number $j$
\begin{align}
j=1+\sum_i a_i 2^{n-i}.
\end{align}
For example, when $n=2$, we have 
\begin{align}
f^{(0)}_1=f^{(0)}_{0,0}~,~~f^{(0)}_2=f^{(0)}_{0,1} ~,~~f^{(0)}_3=f^{(0)}_{1,0}~,~~f^{(0)}_4=f^{(0)}_{1,1}  \, .
\end{align} 

It can be derived from \eqref{eq:ibpdt} that  all IBP relations $eq[\nu_0,\bm{a}]$  corresponding to a selected $\nu_0$ can be expressed in the form
\bea \left(M_1\right)_{\bm{b}\bm{a}} f^{(-1)}_{\bm{a}}+\left(M_0\right)_{\bm{b}\bm{a}} f^{(0)}_{\bm{a}}=0 \label{eq:matrixindxibpn}\eea
where the matrix elements are given by 
\bea \left(M_1\right)_{\bm{b}\bm{a}} & = &  \sum_{j=1}^n \left[ \left(M_{1}^{(j)}\right)_{b_j a_j} \prod_{i\neq j}\delta_{b_i a_i} \right] + \nu_0 \delta_{\bm{b}\bm{a}}  \nn\\
\left(M_0\right)_{\bm{b}\bm{a}} &= &  \sum_{j=1}^n \left[ \left(M_{0}^{(j)}\right)_{b_j a_j} \prod_{i\neq j}\delta_{b_i a_i} \right] + ik_0 \delta_{\bm{b}\bm{a}}~~\label{eq:matrixibpn-1}\eea
with $M_{0}^{(j)},M_{1}^{(j)}$ given in \eqref{eq:matrixibp}. The matrix can be
compactly represented as 
\bea  M_1 &= & \sum_{j=1}^n \left(\nu_j+\frac{1}{2}\right)\Lambda_3^{(j)} + \left(\nu_0-\frac{n}{2}-\sum_{i=1}^n \nu_i\right)\mathbb{I}_{2^n}  \nn  \\
M_0 &= &-i \sum_{j=1}^n k_j \Lambda_2^{(j)} + i k_0 \mathbb{I}_{2^n} \label{eq:matrixibpn-2} \eea
where
\bea \left(\Lambda_k^{(j)}\right)_{\bm{b} \bm{a}}\equiv \left(\sigma_k\right)_{b_j,a_j} \prod_{i\neq j}\delta_{b_i,a_i} , k=1,2,3 \label{eq:LambdaMatrix} \eea
is direct product of a series $2\times 2$ identity matrices except the $j$-th one as Pauli $\sigma_k$ matrix.


From the representation \eqref{eq:matrixibpn-2}, one can see $M_1$ is a diagonal matrix
\begin{align}
&\left(M_1\right)_{\bm{b}\bm{a}}= \begin{cases}
\nu_0-\sum_i a_i(2\nu_i+1) , & \bm{b}=\bm{a} \\
0, & \bm{b}\neq \bm{a}
\end{cases} 
\end{align} 
Here we show IBP relation $eq[\nu_0,\bm{a}]$ for $n=2$ and all $\bm{a}$ in matrix form as example
\begin{align}
&\Big(M_1 \Big| M_0 \Big).\bm{f}^{\mathsf{T}}\nn\\
&=\left(
\begin{array}{cccc|cccc}
 \nu _0 & 0 & 0 & 0 & i k_0 & -k_2 & -k_1 & 0 \\
 0 & \nu _0-2 \nu _2-1 & 0 & 0 & k_2 & i k_0 & 0 & -k_1 \\
 0 & 0 & \nu _0-2 \nu _1-1 & 0 & k_1 & 0 & i k_0 & -k_2 \\
 0 & 0 & 0 & \nu _0-2 \nu _1-2 \nu _2-2 & 0 & k_1 & k_2 & i k_0 \\
\end{array}
\right).\bm{f}^{\mathsf{T}}=0\, ,\nn\\
&~~\bm{f}=\{\bm{f}^{(-1)},\bm{f}^{(0)}\},~~~~~\bm{f}^{(i)}=\{f^{(i)}_{0,0},f^{(i)}_{0,1},f^{(i)}_{1,0},f^{(i)}_{1,1}\}\, ,
\end{align} 
where 
\begin{align}
&M_1=M^{(1)}_1 \otimes \mathbb{I}_2 + \mathbb{I}_2 \otimes M^{(2)}_1 + \nu_0 \mathbb{I}_2 \otimes \mathbb{I}_2 \nn\\
&M_0=M^{(1)}_0 \otimes \mathbb{I}_2 + \mathbb{I}_2 \otimes M^{(2)}_0 + i k_0 \mathbb{I}_2 \otimes \mathbb{I}_2.
\end{align}
Obviously, 
\begin{align}
A_-(\nu_0)=-M_1^{-1}.M_0\, , ~~~~A_+(\nu_0-1)=-M_0^{-1}.M_1\,.
\end{align}

In above expression, the inverse of diagonal matrix $M_1$ is trivial, but
the inverse of  $M_0$ is non-trivial. Thus we are looking for 
different representation.  Noticing that $M_0$ is formed by the direct product of $\mathbb{I}_2$ and $\sigma_2$, diagonalizing  $\sigma_2$ by redefining each $h(\nu_i,a_i;-k\tau)$ immediately leads to the diagonalization of $M_0$:
\bea \tilde{h}(\nu_i,a_i;-k\tau) = \sum_{b_i=0,1} \text{T}_{a_i b_i} h(\nu_i,b_i;-k\tau)\eea
with
\bea \text{T}=\frac{1}{\sqrt{2}}\left(
\begin{array}{cc}
 1 & -i \\
 -i & 1 \\
\end{array}
\right)\, , ~~\text{T}^{-1}=\frac{1}{\sqrt{2}}\left(
\begin{array}{cc}
 1 & i \\
 i & 1 \\
\end{array}
\right)  \label{eq:T}\eea
%
This leads to
\begin{align}
&\tilde{V}(\nu_0,\bm{a})=\int_{-\infty}^0 \tau^{\nu_0} e^{ik_0 \tau} 
  \prod_i \tilde{h}(\nu_i,a_i;-k_i\tau)\nd \tau \nn\\
&\tilde{\bm{f}}^{(a_0)}=\text{T}_n.\bm{f}^{(a_0)}\, , ~~~~~\left(\text{T}_n\right)_{\bm{b}\bm{a}}=\prod_{i=1}^n T_{b_ia_i}\, ,~~~\left(\text{T}_n^{-1}\right)_{\bm{b}\bm{a}}=\prod_{i=1}^n T^{-1}_{b_ia_i}\, \label{eq:Ta}
\end{align}
Using the properties
\begin{align}
&\text{T}.\sigma_2.\text{T}^{-1}=\sigma_3\,,~~~\text{T}.\sigma_3.\text{T}^{-1}=-\sigma_2 \, , ~~~ \text{T}.\sigma_1.\text{T}^{-1}=\sigma_1 \, , ~~~ \text{T}.\mathbb{I}_2.\text{T}^{-1}=\mathbb{I}_2 \, , \nn\\
&\text{T}_n.\Lambda_2^{(j)}.\text{T}^{-1}_n=\Lambda_3^{(j)}\, ,~~~~\text{T}_n.\Lambda_3^{(j)}.\text{T}^{-1}_n=-\Lambda_2^{(j)} \, .   \label{eq:Tsigmai} 
\end{align}
after applying the transformation \eqref{eq:T} and \eqref{eq:Ta}  to \eqref{eq:matrixibp}, we have
\begin{align}
&\tilde{M}_{1}^{(j)}=-\frac{2\nu_i+1}{2}\left(\mathbb{I}_{2}+\sigma_2 \right) \, , ~~~~ \tilde{M}_{0}^{(j)}= - i k_j \sigma_3 \,, \nn\\ 
&\tilde{M}_1= - \sum_{j=1}^n \left(\nu_j+\frac{1}{2}\right) \Lambda_2^{(j)}   + \left(\nu_0-\frac{n}{2}-\sum_{i=1}^n \nu_i\right) \mathbb{I}_{2^n} , \nn\\
&\tilde{M}_0= - \sum_{j=1}^n i k_j \Lambda_3^{(j)} + i k_0 \mathbb{I}_{2^n}.\label{eq:matrixibpntilde}
\end{align} 
Now, on the contrary,  $\tilde{M}_0$ is a diagonal matrix
\begin{align}
&\left(\tilde{M}_0\right)_{\bm{b}\bm{a}}= \begin{cases}
i \left( k_0+\sum_{i=1}^n (2a_i-1)k_i\right) , & \bm{b}=\bm{a} \\
0, & \bm{b}\neq \bm{a}
\end{cases}
\end{align} 
while $\tilde{M}_1$ is not. For example, when $n=2$, we have 
\begin{align}
&\Big(\tilde{M}_1 \Big| \tilde{M}_0 \Big).\bm{\tilde{f}}^{\mathsf{T}}=0\, ,~~~~~\bm{\tilde{f}}=\{\bm{\tilde{f}}^{(-1)},\bm{\tilde{f}}^{(0)}\} \, , \nn\\
&\tilde{M}_1=\left(
\begin{array}{cccc}
 \nu _0-\nu _1-\nu _2-1 & \frac{1}{2} i \left(2 \nu _2+1\right) & \frac{1}{2} i \left(2 \nu _1+1\right) & 0 \\
 -\frac{1}{2} i \left(2 \nu _2+1\right) & \nu _0-\nu _1-\nu _2-1 & 0 & \frac{1}{2} i \left(2 \nu _1+1\right) \\
 -\frac{1}{2} i \left(2 \nu _1+1\right) & 0 & \nu _0-\nu _1-\nu _2-1 & \frac{1}{2} i \left(2 \nu _2+1\right) \\
 0 & -\frac{1}{2} i \left(2 \nu _1+1\right) & -\frac{1}{2} i \left(2 \nu _2+1\right) & \nu _0-\nu _1-\nu _2-1 \\
\end{array}
\right) \nn\\
&\tilde{M}_0= i \left(
\begin{array}{cccc}
 k_0-k_1-k_2 & 0 & 0 & 0 \\
 0 & k_0-k_1+k_2 & 0 & 0 \\
 0 & 0 & k_0+k_1-k_2 & 0 \\
 0 & 0 & 0 & k_0+k_1+k_2 \\
\end{array}
\right)
\end{align} 

Putting all together finally we have
\begin{align}
&A_-(\nu_0)=-M_1^{-1}.M_0\, , \nn\\
&A_+(\nu_0-1)=-\text{T}_n^{-1}.\tilde{M}_0^{-1}.\tilde{M}_1.\text{T}_n=-\text{T}_n^{-1}.\tilde{M}_0^{-1}.\text{T}_n. M_1\,, \label{eq:nibp}
\end{align}
where  $M_1^{-1}$, $\tilde{M}_0^{-1}$ and their inverse are diagonal matrices, with $M_1$ merely relying on $\nu_i$, $\tilde{M}_0$ only relying on $k_i$ and $\text{T}_n$  a constant matrix.

Before ending this subsection, let us give a remark.
Let's see what will happen if we haven't asked the definition of $h(\mu,\bm{a})$ to cancel the $1/\tau^2$ in \eqref{eq:eom1}. For example, define $h(\nu_0,\bm{a})=(-k\tau)^{\frac{1}{2}}\text{H}_\nu^{(1,2)}$ in \eqref{eq:vtxfamily} which satisfy
\begin{align}
\partial_{\tilde{\tau}}^2 h(\mu,0;\tilde{\tau})+\Big(1+\frac{\mu^2}{\tilde{\tau}^2}\Big)h(\mu,0;\tilde{\tau})=0,~~~\tilde{\tau}=-k\tau,~~\mu^2=\frac{m^2}{H^2}-2
\end{align}
we will have 
\begin{align}
&M_{2}^{(1)}.\bm{f}^{(-2)}+\nu_0\mathbb{I}_{2}.\bm{f}^{(-1)}+\left(M_{0}^{(1)}+ik_0\mathbb{I}_{2}\right)).\bm{f}^{(0)}\nn\\
&M_{2}^{(j)}=\left(
\begin{array}{cc}
 0 & 0 \\
 \frac{\mu_j^2}{k_j} & 0 \\
\end{array}
\right) , ~~~~ M_{0}^{(j)}=\left(
\begin{array}{cc}
 0 & -k_j \\
 k_j & 0 \\
\end{array}
\right). 
\end{align} 
The formula of reduction will not be as explicit as it is now due to the emergence of $\bm{f}^{(-2)}$ from the $1/\tau^2$ term. That's why we request this term to be canceled in the definition.

\subsection{n-fold vertex integral family: universal formula of $\nd \log$-form DE}  \label{sec:3.5}
For n-fold vertex integral family, with \eqref{eq:ibpdk}, acting $\partial_{k_0}$ and $\partial_{k_i}$ on $\bm{f}^{(0)}$ give
\begin{align}
&\partial_{k_0}\bm{f}^{(0)}=i \bm{f}^{(1)} = i A_+(\nu_0)\bm{f}^{(0)} \,, \nn\\
&\partial_{k_i}\bm{f}^{(0)} = \left( - \frac{1}{ k_i}\frac{2\nu_i+1}{2}(\mathbb{I}_{2^n} - \Lambda_3^{(i)})-i  \Lambda_2^{(i)}.A_+(\nu_0) \right).\bm{f}^{(0)}\, , ~\text{for $i>0$}  \, .  \label{eq:nDE}
\end{align}

Defining
\begin{align}
\left( \tilde{\Omega}_0 \right)_{\bm{b}\bm{a}} & \equiv \begin{cases} - i \log \Big[ k_0+\sum_i (2a_i-1)k_i\Big] , & \bm{b}=\bm{a} \\
0, & \bm{b}\neq \bm{a} 
\end{cases} \, , \nn\\
\left( \Omega_{ex} \right)_{\bm{b}\bm{a}} & \equiv \begin{cases} - \sum_i a_i(2\nu_i+1) \log k_i , & \bm{b}=\bm{a} \\
0, & \bm{b}\neq \bm{a} 
\end{cases} \, ,
\end{align}
we have $\nd \log$-form DE 
\begin{align}
&\nd \bm{f}^{(0)} = (\nd \Omega) . \bm{f}^{(0)} = \sum_{i=0}^{n} \Omega_{k_i} . \bm{f}^{(0)} \nd k_i  \, ,\nn\\
&\Omega =  \Omega_{ex} - i \text{T}_n^{-1}.  \tilde{\Omega}_0.\text{T}_n.M_1(\nu_0+1) \, ,  \label{eq:ndlogDE}
\end{align}
where $M_1(\nu_0+1)$ is shifting the $\nu_0$ in the original $M_1$ to $\nu_0+1$.

Proof:
\begin{align}
& \partial_{k_0} \Omega =   -i \text{T}_n^{-1}.  \partial_{k_0}\tilde{\Omega}_0.\text{T}_n.M_1(\nu_0+1) =  -i \text{T}_n^{-1}.  \tilde{M}_0^{-1} .\text{T}_n.M_1(\nu_0+1) = i A_+(\nu_0) \nn\\
& \partial_{k_i} \Omega = \partial_{k_i}\Omega_{ex} - i \text{T}_n^{-1}.  \partial_{k_i}\tilde{\Omega}_0.\text{T}_n.M_1(\nu_0+1) \nn\\
&~~~~~~=- \frac{1}{ k_i}\frac{2\nu_i+1}{2}(\mathbb{I}_{2^n} - \Lambda_3^{(i)}) + i \text{T}_n^{-1}.  \Lambda_3^{(i)}.\tilde{\Omega}_0.\text{T}_n.M_1(\nu_0+1) \nn\\
&~~~~~~=- \frac{1}{ k_i}\frac{2\nu_i+1}{2}(\mathbb{I}_{2^n} - \Lambda_3^{(i)}) + i  \Lambda_2^{(i)}. \text{T}_n^{-1}.  \tilde{\Omega}_0.\text{T}_n.M_1(\nu_0+1)   \nn\\
&~~~~~~=- \frac{1}{ k_i}\frac{2\nu_i+1}{2}(\mathbb{I}_{2^n} - \Lambda_3^{(i)}) -i  \Lambda_2^{(i)}.A_+(\nu_0)  \, .
\end{align}
Comparing these equations with \eqref{eq:nDE}, the proof is complete.

Then, once people have got boundary condition of $\bm{f}^{(0)}$, $\bm{f}^{(0)}(k_0^{0},k_1^0,\cdots)$ for example, $\bm{f}^{(0)}(k_0',k_1',\cdots)$ can be got by
\begin{align}
&\bm{f}^{(0)}(k_0',k_1',\cdots) = \bm{f}^{(0)}(k_0^{0},k_1^0,\cdots) \nn\\
 & \quad \quad \quad + \mathcal{P} \exp\left[\int_{(k_0^{0},k_1^0,\cdots)}^{(k_0',k_1',\cdots)}  \sum_i \Omega_{k_i}(\nu_0,\nu_1,\cdots;k_0,k_1,\cdots) \nd k_i  \right]  .\ \bm{f}^{(0)}(k_0^{0},k_1^0,\cdots)
\end{align}
for a integration path starts from $(k_1^{0},k_2^0,\cdots)$ and end at $(k_1',k_2',\cdots)$. $\mathcal{P}$ for path ordering.

At loop-level, these differential equations with respect to loop momentum will also directly give the $\nd\tau_i$-reduced IBP relations we have mentioned near the end of section \ref{sec:3.1}.


\subsection{0-fold vertex integral family: massless cases}  \label{sec:3.6}
For a vertex only connect to massless propagator, the vertex function family is given by
\begin{align}
V(\nu_0)=\int \tau^{\nu_0} e^{i k_0 \tau}.
\end{align}
It has only $2^0=1$ MI and the iterative reduction relations are given by
\begin{align}
&eq[\nu_0]: i k_0 V(\nu_0) + \nu_0 V(\nu_0-1) = 0, \nn\\
&A_+(\nu_0) = i\frac{\nu_0+1}{k_0}\,,~~~A_-(\nu_0) = -i\frac{k_0}{\nu_0}\,. 
\end{align}
and $\nd \log$-form DE is given by
\begin{align}
&\partial_{k_0} V(\nu_0) = i V(\nu_0+1) = -(\nu_0+1) \frac{1}{k_0} V(\nu_0)  \, , \nn\\
&\nd \Omega = \Omega_{k_0} \nd k_0 = -(\nu_0+1) \nd\log(k_0) \, .
\end{align}

\subsection{Discussion on remaining terms come from $G_{\pm\pm}$} \label{sec:3.7}
\subsubsection{Iterative reduction}
Consider an simple example with remaining terms. Suppose $f^{(a_0)}_{\bm{a};1}$ are integrals belong to vertex integral family $V_1$, corresponding to the integration of $\nd \tau_1$, and  $f^{(b_0)}_{\bm{b};2}$ are integrals belong to vertex integral family $V_2$. We use $\hat{f}$ to denote their integrand. For example
\begin{align}
\hat{f}^{(a_0)}_{\bm{a};1} = e^{i k_{0;1}\tau_1} \tau_1^{\nu_{0;1}+a_0} \prod_i   h(\nu_{i;1},a_i,-k_{i;1}\tau_1) \,.
\end{align}
We also denote the combination of two terms in propagator as
\begin{align}
&h(\nu_{i;1},a_i,-k_{i;1}\tau_1) \theta_{1,2}^{(i,j)} h(\nu_{j;2},b_j,-k_{j;2}\tau_2)\equiv  \nn\\
&h^{(1)}(\nu_{i;1},a_i,-k_{i;1}\tau_1) \theta_{12} h^{(2)}(\nu_{j;2},b_j,-k_{j;2}\tau_2) + h^{(2)}(\nu_{i;1},a_i,-k_{i;1}\tau_1) \theta_{21} h^{(1)}(\nu_{j;2},b_j,-k_{j;2}\tau_2) \, , \nn\\
&\nu_{i;1}= \nu_{j;2} \,, \ \ k_{i;1}=k_{j;2} \, .
\end{align}
Then, if there is a $G_{++}$, we denote the integrals as
\begin{align}
\bm{f}^{(a_0,b_0)}_{\bm{a},\bm{b}} \equiv \int \nd \tau_1 \nd \tau_2  \hat{f}^{(a_0)}_{\bm{a};1} \theta_{1,2}^{(i,j)} \hat{f}^{(b_0)}_{\bm{b};2} \label{eq:egR}
\end{align}
Without loss of generality, consider IBP of $\nd \tau_1$ and $a_0=b_0=0$, one could have IBP relations
\begin{align}
 &\left(M_1\right)_{\bm{c}\bm{a}} \bm{f}^{(-1,0)}_{\bm{a},\bm{b}}+\left(M_0\right)_{\bm{c}\bm{a}} \bm{f}^{(0,0)}_{\bm{a},\bm{b}} + \delta_{\bm{c} \bm{a}} \bm{R}_{\bm{a},\bm{b}}^{(0,0)} =0 \, , 
\end{align}
which is just adding remaining terms of sub-sector and the factor $\theta_{1,2}^{(i,j)} \hat{f}^{(0)}_{\bm{b};2}$ to \eqref{eq:matrixindxibpn}. According to \eqref{eq:vanishR} and \eqref{eq:pinchR}, 
\begin{align}
&\bm{R}_{\bm{a},\bm{b}}^{(a_0,b_0)}=\delta_{a_i,1-b_j} (-1)^{a_i+1} \frac{4i}{\pi} e^{ \pi \text{Im}[\nu]} (-k_{i;1})^{-2\nu_{i;1}-1} f^{(a_0+b_0-2\nu_{i;1}-1)}_{\bm{a}_{\hat{i}},\bm{b}_{\hat{j}} } \,, \nn\\
&\bm{a}_{\hat{i}},\bm{b}_{\hat{j}} = a_1,a_2,\cdots,a_{i-1},a_{i+1},\cdots,a_{n_1},b_1,\cdots, b_{j-1},b_{j+1},\cdots, b_{n_2}  \, ,    \label{eq:Rterm}
\end{align}
where $f^{(a_0+b_0-2\nu_{i;1}-1)}_{\bm{a}_{\hat{i}},\bm{b}_{\hat{j}} }$ is integral belong to $(n_1+n_2-2)$-fold vertex integral family.
Then, reduction of these sector with two vertex integral family could be achieved via using 
\begin{align}
\bm{f}^{(-1,0)}_{\bm{c},\bm{b}}
&= \left(A_{-;1}(\nu_0)\right)_{\bm{c}\bm{a}} \bm{f}^{(0,0)}_{\bm{a},\bm{b}} 
- \left(M_{1;1}^{-1}\right)_{\bm{c}\bm{a}}\bm{R}^{(0,0)}_{\bm{a},\bm{b}}   \, ,\nn\\
\bm{f}^{(1,0)}_{\bm{c},\bm{b}}
&=\left( A_{+;1}(\nu_0)\right)_{\bm{c}\bm{a}}  \bm{f}^{(0,0)}_{\bm{a},\bm{b}}
-\left(\text{T}_{n_1}^{-1}.\tilde{M}_{0;1}^{-1}.\text{T}_{n_1}\right)_{\bm{c}\bm{a}}\bm{R}^{(1,0)}_{\bm{a},\bm{b}} \, \label{eq:ibpwithR}
\end{align}
iteratively. In these equations, the subscript $;1$ means they are the expression with respect to family $V_1$. In other words, the $A_{\pm;1}$, $M_{1;1}$, and $\tilde{M}_{0;1}$ here are just given by the same formulas as \eqref{eq:nibp}, \eqref{eq:matrixibpn-2}, and \eqref{eq:matrixibpntilde}. These formulas give the iterative reduction relations without computing  inverse of large matrix. For more than one $G_{\pm\pm}$, the reduction is similar to \eqref{eq:ibpwithR}, but with more remaining terms corresponding to each $G_{\pm\pm}$. Since all remaining terms are also product of vertex integral families, one could reduce them via formula like \eqref{eq:ibpwithR} as well, thus the reduction is complete.
We have got the universal formula for reducing all tree-level dS integrals.

\subsubsection{$\nd \log$-form DE}
Consider the example \eqref{eq:egR} again, DE is given by
\begin{align}
&\left(\partial_{k_{0;1}}\bm{f}^{(0,0)}\right)_{\bm{c},\bm{b}}=i \bm{f}^{(1,0)}_{\bm{c},\bm{b}}   \,, \nn\\
&\left(\partial_{k_{i;1}}\bm{f}^{(0,0)}\right)_{\bm{c},\bm{b}} =  - \frac{1}{ k_i}\frac{2\nu_i+1}{2}\left(\mathbb{I}_{2^n} - \Lambda_3^{(i)}\right)_{\bm{c}\bm{a}}\bm{f}^{(0,0)}_{\bm{a},\bm{b}}-i  \left(\Lambda_2^{(i)}\right)_{\bm{c}\bm{a}} \bm{f}^{(1,0)}_{\bm{a},\bm{b}}\, , ~\text{for $i>0$}  \, ,  \label{eq:nDE11}
\end{align}
where the $\bm{f}^{(1,0)}_{\bm{a},\bm{b}}$ is reduced  by \eqref{eq:ibpwithR}. As expected, the additional term arises from the remaining terms $\bm{R}_{\bm{a},\bm{b}}^{(0,0)}$ that emerge during its reduction. Obviously, if we choose these $\frac{4i}{\pi} e^{ \pi \text{Im}[\nu]} (-k_{i;1})^{-2\nu_{i;1}-1} f^{(-2\nu_{i;1})}_{\bm{a}_{\hat{i}},\bm{b}_{\hat{j}} }$ in remaining terms \eqref{eq:Rterm}  to be master integrals of this sub-sector, the dependence of the original sector on  the sub-sector in the differential equation matrix is
\begin{align}
 &\Omega_{({\bm{a},\bm{b}})(\bm{c}_{\hat{i}},\bm{d}_{\hat{j}})} =- i  \sum_{c_i}\left( \text{T}_n^{-1}.  \tilde{\Omega}_{0;1}.\text{T}_n \right)_{\bm{a}\bm{c}} \delta_{\bm{b}\bm{d}} \delta_{c_i(1-d_j)} (-1)^{c_i+1} \nn\\
&~~~~~~~~~~~~~~~~~ - i \sum_{d_j} \left( \text{T}_n^{-1}.  \tilde{\Omega}_{0;2}.\text{T}_n \right)_{\bm{b}\bm{d}} \delta_{\bm{a}\bm{c}} \delta_{d_j(1-c_i)}  (-1)^{d_j+1} \nn\\
&~~~~~~~~~~~~~~~ =- i \left( \text{T}_n^{-1}.  \tilde{\Omega}_{0;1}.\text{T}_n \right)_{\bm{a}(\bm{c}_{\hat i};1-b_j)} \delta_{b_{\hat j}d_{\hat j}} (-1)^{b_j}  \nn\\
&~~~~~~~~~~~~~~~~~~ -  i \left( \text{T}_n^{-1}.  \tilde{\Omega}_{0;2}.\text{T}_n \right)_{\bm{b}(\bm{d}_{\hat i};1-a_i)} \delta_{a_{\hat i}c_{\hat i}} (-1)^{a_i}
\end{align}  
where for the subsript of $\Omega_{({\bm{a},\bm{b}})(\bm{c}_{\hat{i}},\bm{d}_{\hat{j}})}$,  ${\bm{a},\bm{b}}$ are the indices for original sector and $\bm{c}_{\hat{i}},\bm{d}_{\hat{j}}$ are indices for the sub-sector.
Then, it is again $\nd \log$-form DE. Also notice that $\nu_{i;1}=\nu_{j;2}, \ k_{i;1}=k_{j;2}$, so the master integral $\frac{4i}{\pi}  e^{ \pi \text{Im}[\nu]} (-k_{i;1})^{-2\nu_{i;1}-1} f^{(-2\nu_{i;1})}_{\bm{a}_{\hat{i}},\bm{b}_{\hat{j}} }$  will be consistent for $;1$ and $;2$. For more than one $G_{\pm\pm}$, the discussion is similar.


\section{Summary and Outlook}\label{sec:4}
In this paper, we define the dS integral family, which naturally incorporates the case of time derivatives interaction. We generalize IBP method to dS integral family with respect to $\nd \tau_i$ and $\nd k_i$ by \eqref{eq:ibpdt} and \eqref{eq:ibpdk}, whose integrands involve special functions. With \eqref{eq:ibpdk} and result of IBP reduction,  people also can construct DE with respect to $\partial_{k_i}$ of dS integrals. We indicate  the factorization of IBP relations of vertex in the dS correlator. For the vertex integral families, we derive an universal iterative reduction formula for arbitrary $n$-fold Hankel vertex function families, along with the $\nd \log$-form DE satisfied by the MI we selected, as listed in \eqref{eq:nibp} and \eqref{eq:ndlogDE}. And the remaining terms come from $G_{\pm\pm}$ are also discussed.  Since the tree-level dS correlators only involve integrals over $\tau_i$, we have obtained the reduction and DE of arbitrary tree-level diagrams 
equivalently.

This paper in fact has presented an alternative pathway toward systematically and efficiently computing dS correlators. Once we have IBP relations, drawing from the experience in flat spacetime, the number of integrals people need to compute will be significantly reduced. Once we have DE, with proper boundary conditions, the remaining steps do not differ from the case of flat spacetime, and numerical result can be efficiently obtained via numerical DE method.  Its effectiveness has been validated in flat spacetime, and many existing packages designed for flat spacetime can also be readily applied. For example, Kira could solve IBP reduction of user-defined system \cite{Maierhofer:2018gpa}, AMFlow \cite{Liu:2022chg} and DiffExp \cite{Hidding:2020ytt} can numerically solving DE with given boundary condition and DE.

This paper also suggest many interesting open questions. We merely list a part of them as follows to inspire people's future research. Firstly, while we mainly focus on the IBP linear system, how to give a boundary condition of DE is beyond the scope of this paper and haven't been discussed. One can work like the flat space DE, select a simple boundary and analytically or numerically compute it. But whether there is a  boundary-determine method like AMFlow 
 \cite{Liu:2017jxz,Liu:2022mfb} in flat cases needs more consideration. For example, one may consider the boundary $k_0\to -i \infty$ in dS cases, which shrink all vertex integrals to zero. Secondly, AdS correlator could be considered in the future and the relation of dS and AdS integral could be examined in the perspective of IBP and DE. Thirdly, follow the idea that reclassifying Feynman integrals itself as special function via DE \cite{Liu:2023jkr},
to bring dS integrals closer to a so called analytical result, people need to develop systematical method to analyze this integral family. For example, people may consider how to distinguish the different signal and background in the dS correlator using DE formalism like people have done  using partial MB transformation \cite{Qin:2022fbv} in the future.

\section*{Acknowledgements}
This work is supported by Chinese NSF funding under Grant No.11935013, No.11947301, No.12047502 (Peng Huanwu Center), No.12247120, No.12247103, No.U2230402, and China Postdoctoral Science Foundation No.2022M720386.

\bibliographystyle{JHEP}
\bibliography{references}

\end{document}